\documentclass[prd,
eqsecnum,floatfix,letterpaper,showpacs,superscriptaddress,groupedaddress,nofootinbib]{revtex4-2}

\def \vss{\vspace{14pt}}
\def \vsss{\vspace{10pt}}

\usepackage{amsmath}
\usepackage{latexsym}
\usepackage{amssymb}
\usepackage{amsfonts}
\usepackage{mathtools}
\usepackage{bm}
\usepackage{amsthm}
\usepackage{graphicx}
\usepackage{color}
\usepackage{changes}
\usepackage[normalem]{ulem}
\usepackage{natbib}
\usepackage{appendix}
\def \be{\begin{equation}}
\def \bea{\begin{eqnarray}}
\def \eea{\end{eqnarray}}
\def \ee{\end{equation}}
\def \no {\nonumber}

\def \D{{\cal D}}
\def \G{{\cal G}}
\def \R{\mathcal{R}}

\def \W{\mathcal{W}}
\def \V{\mathcal{V}}
\def \VG{{\mathcal V}_{\G}}
\def \Vp{{\mathcal V}_\perp}
\def \S{{\cal S}}
\def \P{{\cal P}}

\def \Sgm{\Sigma}
\def \Msun {M_\odot}
\def \bM{{\bf M}}

\def \Ub{{\bf U}}
\def \Vb{{\bf V}}
\def \a{\alpha}
\def \b{\beta}
\def \M{{\mathcal M}}
\def \Mc{{\M_{\rm chirp}}}
\def \fl{f_{\rm lower}}
\def \fu{f_{\rm upper}}
\def \eps{\epsilon}
\def \x{{\bf x}}
\def \y{{\bf y}}
\def \bh{{\bf h}}
\def \Dh{{\bf \Delta h}}
\def \n{{\bf n}}
\def \bg{{\bf g}}
\def \s{{\bf s}}

\def \th{{\tilde h}}

\def \e{{\bf e}}
\def \v{{\bf v}}

\begin{document}
\title{Unified $\chi^2$ discriminators for gravitational wave searches from compact coalescing binaries}

\author{Sanjeev Dhurandhar}
\email{sanjeev@iucaa.in}
\affiliation{Inter University Centre for Astronomy and Astrophysics, Ganeshkhind, Pune, 411 007, India}

\date{\today}

\begin{abstract}
Gravitational wave (GW) signals of astrophysical origin are typically weak. This is because gravity is a weak force, weakest among the four forces we know of. In order to detect GW signals, one must make differential measurements of effective lengths less than thousand'th of the size of a proton. In spite of the detectors achieving extraordinary sensitivity, the detector noise typically overwhelms the signal, so that GW signals are deeply buried in the data. The challenge to the data analyst is of extracting the GW signal from the noise, that is, first deciding whether a signal is present or not then if present, measuring its parameters. 
\vss

However, in the search for coalescing compact binary (CBC) signals, short-duration non-Gaussian noise transients (glitches) in the detector data significantly affect the search sensitivity. Chi-squared discriminators are therefore employed to mitigate their effect. We show that the underlying mathematical structure of any $\chi^2$ is a vector bundle over the signal manifold $\P$, that is, the manifold traced out by the signal waveforms in the Hilbert space of data segments $\D$. The $\chi^2$ is then defined as the square of the $L_2$ norm of the data vector projected onto a finite-dimensional subspace $\S$ (fibre) of $\D$ chosen orthogonal to the triggered template waveform. Any such fibre leads to a $\chi^2$ discriminator and the full vector bundle comprising the subspaces $\S$ and the base manifold $\P$ constitute the $\chi^2$ discriminator. We show that this structure paves the way for constructing effective $\chi^2$ discriminators against different morphologies of glitches. Here we specifically demonstrate our method on blip glitches, which can be modelled as sine-Gausians, which then generates an optimal $\chi^2$ statistic for blip glitches.
\end{abstract}
\maketitle

\section{Introduction}

A spectacular prediction from Einstein's general theory of relativity is the existence of gravitational waves \citep{Einstein1918}. However, gravitational waves (GW) are inherently weak because gravity is a weak force, the weakest among the four known forces. Minute differential interferometric path length measurements must be made to observe these waves with sufficient confidence. During the past several decades and in the more recent past, breakthroughs in technology have made it possible to build kilometre scale laser interferometric detectors on ground.  Past six decades of heroic
experimental efforts undertaken by physicists all over the world have
finally culminated with the first direct observation of a GW signal
announced by the Laser Interferometer Gravitational Wave Observatory
(LIGO) project \citep{LIGO,GW150914}. On September 14, 2015, the two
LIGO interferometers at Hanford (Washington) and Livingston
(Louisiana) simultaneously measured and recorded strain data that
indicated the presence of a GW signal emitted by a coalescing binary
a system containing two black holes of masses around 30 times the mass of the Sun. Since then many more detections, as many as $90$, have been made in the science runs O1, O2 and O3 \citep{gwtc3}. Currently, the LIGO \citep{advligo}, the Italian-French VIRGO \citep{AdvVIRGO}, and the Japanese KAGRA \citep{kagra} are in operation, and the O4 run is in progress. The detector on Indian soil, namely, LIGO-India \cite{ligoindia}, is expected to join in these efforts early next decade. Gravitational wave astronomy has truly arrived.
\par

\begin{figure}[htbp]
\begin{center}
\includegraphics[width=0.5\textwidth]{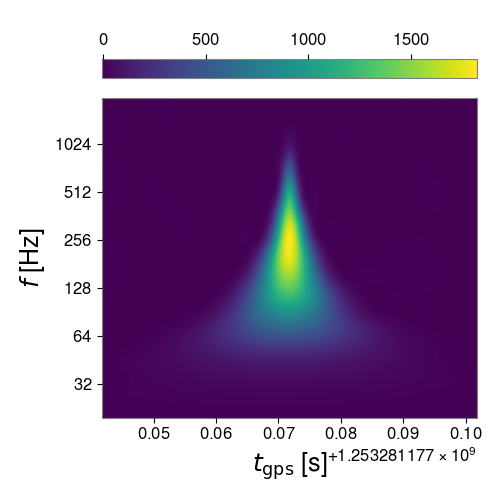}
\caption{The figure shows a typical glitch in the time-frequency plane. This is called a blip glitch because of its distinctive features.}
\label{fig:glitch}
\end{center}
\end{figure}

However, the data in the ground-based detectors is neither Gaussian nor stationary; non-Gaussian and non-stationary features can be produced by various components of the detector itself or by the environment. In this article, we will only consider CBC signals emitted by a pair consisting of either black holes or neutron stars. These signals are essentially transient, lasting between a fraction of a second to several minutes. Further, the signals can be adequately modeled by using either post-Newtonian approximation methods, perturbation techniques, or numerical relativity. In the case of adequately modelled signals, the preferred data analysis technique is matched filtering. However, the signals depend on several parameters, such as the masses, time of coalescence, etc., and then it becomes necessary to employ a bank of templates densely covering the parameter space. In practice, however, matched filtering alone is not sufficient to identify a signal because the data contain non-Gaussianities and transient noise artifacts, also termed as glitches. Fig. \ref{fig:glitch} shows a typical example of a glitch that occurs in the data. 
Even if the overlap of a glitch with a template is small, the glitches themselves can have large amplitudes and they can produce sufficiently loud triggers in the filtered output. Then, such triggers have a chance of being misinterpreted as signals. To remedy this situation, $\chi^2$ vetos or  discriminators were proposed. The first $\chi^2$ discriminator \cite{Allen2005} proposed by Bruce Allen, which we call {\it traditional}, is designed to distinguish between a signal and a glitch by producing a high value of the $\chi^2$ statistic if the trigger arises from a glitch, while a low value of $\chi^2$ indicates the presence of a signal. The statistic is based on how the power of the signal is distributed in various frequency bins. This is done by dividing the data into several frequency bins and checking whether this power distribution is consistent with that of the signal. This is why, it is also called the {\it power} $\chi^2$. We briefly review the traditional $\chi^2$ due to Bruce Allen's in section \ref{subsec:trad_chisqr}. However, this is not the only type of $\chi^2$ discriminator that can be constructed; it can be constructed from general considerations \cite{DGGB2017,HF11}; these works show that a plethora of $\chi^2$s can be constructed. We call this  the generic or the unified $\chi^2$. For example, another $\chi^2$ that is also employed is the Sine-Gaussian (SG) $\chi^2$ \cite{Nitz:2017lco}. The generic $\chi^2$ discriminator is discussed in section \ref{sec:generic} and specifically in subsection \ref{subsec:generic_chi2}. We show that the traditional $\chi^2$ can be brought under the common umbrella of the generic $\chi^2$ discriminators. In section \ref{sec:blip} we construct the optimised $\chi^2$ which we call Opt $\chi^2$ in short. In section \ref{sec:results} we compare the different $\chi^2$s. Finally, in section \ref{sec:concl} we conclude. 
\par

In the next section \ref{sec:context}, we discuss the context and some preliminaries.

\section{The context} 
\label{sec:context}

 In this section we describe the matched filtering technique and how it is employed in the detection of CBC signals. We then discuss the statistical method of deciding on detection and significance. The treatment here follows the frequentist approach. Currently, Bayesian methods are in vogue. However, we do not discuss them here. In fact, gravitational wave astronomy is in its infancy, and it seems somewhat premature to choose priors; we have to rely on our knowledge gained from electromagnetic astronomy, and it could give a misleading picture. The gravitational wave universe could be far different from the universe as we know it from other astronomical windows.
 
\subsection{The matched filtering paradigm}

Consider a data train $x(t)$ defined over a time interval $[0, T]$. The data trains form an infinite dimensional vector space, which we call $\D$.  The data trains as vectors in $\D$ will be denoted in boldface $x(t) \longrightarrow \x$. Let $n(t)$ be the noise in the detector, a stochastic process defined over $[0, T]$; we assume that it has an ensemble mean zero and is stationary in the wide sense, that is, its first two moments are stationary - they do not depend on the absolute time. A given noise realisation is a vector $\n \in \D$ - it is a random vector. We denote its power spectral density (PSD) by $S_h (f)$ as is prevalent in the literature; the subscript $h$ has been introduced, indicating the detector's signal channel. The signal is the gravitational wave strain, denoted by $h(t)$. We also state that $S_h (f)$ is a one-sided PSD - the negative frequencies are folded onto positive frequencies - it is defined only for positive frequencies. Now we consider two data trains $\x$ and $\y$; their scalar product is written conveniently in the Fourier domain. Let $\tilde{x}(f)$ and $\tilde{y}(f)$ be the Fourier representations of $\x$ and $\y$, then the scalar product is given by:
\be 
 (\x ,\y) = 4 \Re \int_{\fl}^{\fu}~ df \frac{\tilde{x}^* (f) \tilde{y}(f)}{ S_{h}(f)} \,,
\label{eq:scalar} 
\ee
where the integration is carried out over the band-width $[\fl, \fu]$. This construction makes the space $\D$ into a Hilbert space - a $L_2$ space with measure $d \mu \equiv df / S_h (f)$. Then $\D = L_2([0, T], \mu)$. This scalar product can be used to normalise vectors in $\D$; $\|\x \| = + \sqrt{(\x, \x)}$.
\par

We think of a signal $\s$ as just an amplitude $A$ multiplying a normalized waveform $\bh$, where $\|\bh \| = 1$; thus, $\s = A \bh$. The data vector, which we denote by $\x$, in additive noise is then $\x = \s + \n$ when a signal is present, while in the absence of a signal, it is just noise, i.e., $\x = \n$. The match $c$ (correlation) is the scalar product between the data $\x$ and a (normalized) template  $\bh$, that is, $c = (\x, \bh)$. 
Eq.(\ref{eq:scalar}) of the scalar product turns this operation into a matched filter because,
\be
c = 4 \Re \int_{\fl}^{\fu}~ df~ \tilde{x}^* (f) \tilde{q}(f) \,,
\label{eq:mflt} 
\ee
where $\tilde{q} (f) = \th (f)/S_h (f)$ where we have the usual $L_2$ scalar product in Eq. (\ref{eq:mflt}). The quantity $\tilde{q} (f)$ is in fact the matched filter. We remark that $c$ is a surrogate statistic of the likelihood ratio in Gaussian noise. 
\begin{figure}[htbp]
\begin{center}
\includegraphics[width=0.55\textwidth]{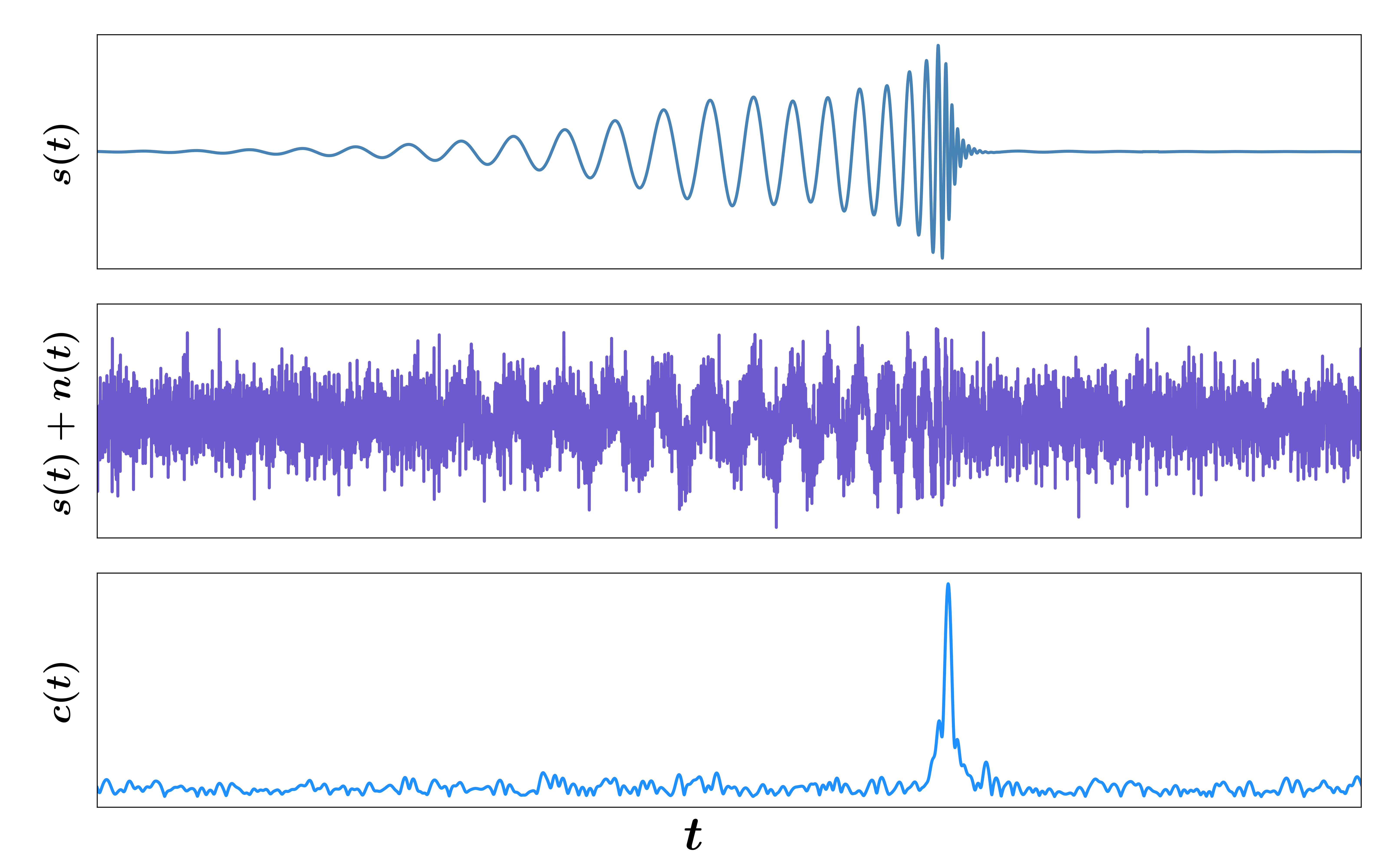}
\caption{The top panel shows the signal $s(t)$, while the middle panel shows the signal embedded in the noise $x(t) = s(t) + n(t)$, where the amplitude of the signal has been reduced, so that it is not apparent in the data. The bottom panel shows the matched filter output $c$ which has a peak, in spite of the weak signal. If the peak crosses a preset threshold, we then say we have detected a signal.}
\label{fig:mflt}
\end{center}
\end{figure} 
Fig. \ref{fig:mflt} displays this operation on a data segment that contains noise plus signal. See also \citep{Dhurandhar-Mitra}. 

\subsection{Maximum likelihood detection}

 First consider the simple situation of a binary hypothesis. Let $p_0 (\x)$ is the probability density function (pdf) when the signal is absent, that is, $\x = \n$ (hypothesis $H_0$) and $p_1 (\x)$ is the pdf when the signal is present, that is, $\x = \s + \n$ (hypothesis $H_1$). Detection is decided by computing the likelihood ratio $\Lambda (\x) = p_1 (\x)/p_0(\x)$ and comparing its maximum  with a threshold $\Lambda_0$, which in turn depends on the false alarm probability $P_F$ that we can tolerate. In order to achieve this, the space $\D$ is partitioned into 
$\D = \R + \R^c$, where the superscript "$c$" denotes the complement. The false alarm and detection probabilities are defined as:
\be
P_F (\R) = \int_{\R} p_0 (\x)~ d \x \,,  ~~~~~~~~~P_D (\R) = \int_{\R} p_1 (\x)~ d \x \,.
\label{eq:PFPD}
\ee
The region $\R$ is found by the Neyman-Pearson lemma \citep{Helstrom}, which maximises the detection probability for a given false alarm probability. The likelihood ratio must satisfy $\Lambda (\x) \geq \Lambda_0$, and $\Lambda_0$ chosen so that $P_F = \a$, where $\a$ is the false alarm probability that we can tolerate, usually $\a <<1$. This sets the threshold $\Lambda_0$. Detection is announced if $\Lambda (\x) > \Lambda_0$.
\par

\begin{figure}[htbp]
\begin{center}
\includegraphics[width=0.4\textwidth]{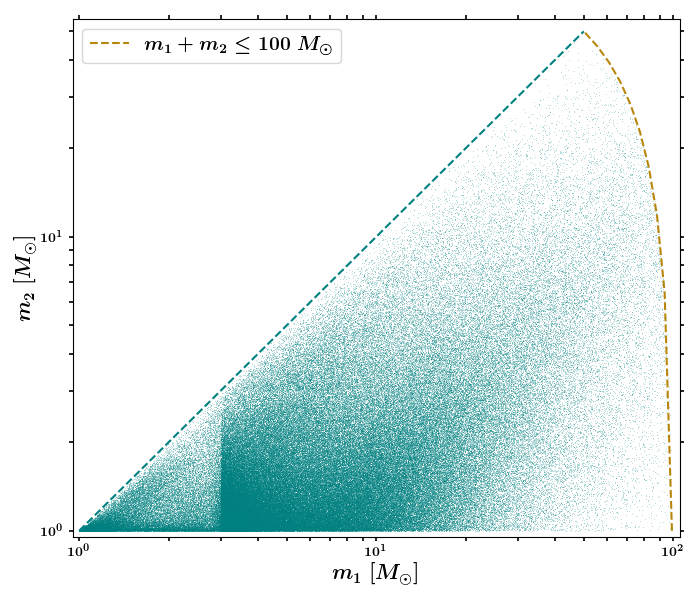}
\caption{In the template bank shown, the ranges of the masses $m_1, m_2$ are as follows: $1 \Msun \leq m_1, m_2 \leq 100 \Msun$, $m_1 + m_2 \leq 100 \Msun$ and $m_2 \leq m_1$. The bank consists of about a million templates.}
\label{fig:template_bank}
\end{center}
\end{figure}

However, we are not in this situation because the signal depends on several parameters; that is, the signal is one among a family of signals, and then we have a composite hypothesis $H_1$. We must, therefore, use maximum likelihood detection. In Gaussian noise, we can replace the likelihood ratio by the surrogate statistic $c$ as it is a monotonic function of $c$. In searching for signals, the statistic $c$ is maximized over template parameters and compared with a threshold based on $P_F$. In the simplest case, the signal is a function of the masses $m_1, m_2$ and kinematical parameters, namely, the coalescence time $t_c$ (in Fig. \ref{fig:mflt} the peak occurs at $t = t_c$) and the coalescence phase $\phi_c$. In practice, for the parameters $t_c, \phi_c$, the templates (normalised waveforms) need to be only defined at $\phi_c = 0$ and $\phi_c = \pi/2$, and for $t_c = 0$. This is because the search over these parameters can be done efficiently using quadratures for $\phi_c$ and the fast Fourier transform algorithm for $t_c$. However, the search over the mass parameters requires a densely sampled discrete bank of templates so that the chance of missing out on a signal is small \citep{SD91,DS94}. A typical template bank is shown in Fig. \ref{fig:template_bank} for masses in the range $1 \Msun \leq m_1, m_2 \leq 100 \Msun$ and $m_1 + m_2 \leq 100 \Msun$.
We only require templates with $m_2 \leq m_1$, because the signal waveform is symmetric in $m_1$ and $m_2$. The density of templates is governed by the maximum mismatch permitted. The match between two normalised templates $\bh_\a$ and $\bh_\b$ is just the scalar product $(\bh_\a, \bh_\b) = \M$ say. The mismatch is defined as $1 - \M$. It is customary to choose the mismatch to be $\sim 3 \%$, which means a possible loss of about 10 \% in the detection of sources as the number of sources depends on the volume (if $\eps$ is the mismatch, then the fraction of signals lost is $1 - (1 - \eps)^3 \simeq 3 \eps$). 

\section{What is a $\chi^2$ discriminator?}
\label{sec:generic}
  As mentioned in the introduction, the data contain non-stationary and non-Gaussian features which are called glitches. These interfere with our maximum likelihood methods; we need to distinguish between triggers produced by signals and those by glitches. For this purpose, we use $\chi^2$ discriminators. We first describe the general mathematical formalism and then use it to target specific morphologies of glitches.  
  
\subsection{Constructing a generic $\chi^2$ discriminator}
\label{subsec:generic_chi2}

We begin with an arbitrary set of linearly independent vectors $\bh_\a \in \D$, where $\a = 1, 2, ...,p$. The $\chi^2$ statistic is constructed by taking the differences between the expected and observed correlations. The expected correlations are obtained when the data are just the signal. The  observed value of the correlation of the data $\x$ with the chosen vector $\bh_{\a}$ is just given by:
\be
c_{\a}^{o} (\x) = (\x, \bh_{\a}) \,
\ee
We now define the expected correlations $c^{e}_{\a}$. We consider the situation of the perfect match between the signal and the normalised trigger template $\bh_0$, and so $\s = A \bh_0$ for some amplitude $A = (\s, \bh_0)$. Then, the correlation of the data $\x$ with the template $\bh_0$ is 
just $c = (\x, \bh_0)$ and its average value is $\langle c \rangle = (\s, \bh_0)$, because $\langle (\n, \bh_0) = 0$. 
We also have the relation, 
\be
(\s, \bh_\a) ~=~ (A \bh_0, \bh_\a) ~=~ A (\bh_0, \bh_\a) ~=~(\s, \bh_0)~(\bh_0, \bh_{\a}) ~=~ \langle c \rangle (\bh_0, \bh_{\a}) \,.
\ee
In the above equation, we replace the average value of the correlation $c$, namely, $\langle c \rangle$ by $c$ itself. This gives us the expected correlation:
\be
c_{\a}^{(e)} (\x) = (\x, \bh_0)~(\bh_0, \bh_{\a}) \,.
\label{eq:expected}
\ee
We take the difference between the expected and observed correlations:
\be
\Delta c_{\a} (\x) =  c_{\a}^{o} (\x) - c^{e}_{\a} (\x) \equiv (\x, \Delta \bh_{\a}) \,,
\label{eq:delc}
\ee
where we have defined the projected vectors $\Delta \bh_{\a}$ as,
\be
\Delta \bh_{\a} = \bh_{\a} - (\bh_\a, \bh_0)~ \bh_0 \,.
\label{eq:dhalpha}
\ee
We readily verify that $(\Delta \bh_{\a}, \bh_0) = 0$ for every $\a$ and therefore $\Delta \bh_{\a}$ are also orthogonal to the signal $\s$. Then $\Dh_\a$ span a $p$-dimesional subspace of $\D$. We call this subspace as $\S$. Then the generic $\chi^2$ for an arbitrary collection of vectors $\bh_\a$ is just the square of the $L_2$ norm of the projected data vector $\x$ on $\S$. 
\par

We note the following properties of the linear functional $\Delta c_{\a}$: 
\be
\Delta c_{\a} (\s)  =  0, ~~~~~~~~ \langle \Delta c_{\a} (\n) \rangle = 0 \,.
\ee  
Secondly, the r.v.s $\Delta c_{\a}$ are Gaussian if the noise is Gaussian because they are linear combinations of Gaussian variables, as seen from Eqs. (\ref{eq:delc}) and (\ref{eq:dhalpha}). However, we cannot right away take the sum of squares of the $\Delta c_{\a}$ to construct a $\chi^2$ because (i) they could, in general, be correlated and, moreover, (ii) they may not have unit variance. But these problems can be easily remedied. To construct the $\chi^2$, we first require the following covariance matrix:
\be
C_{\a \b} \equiv \langle \Delta c_{\a} (\n) \Delta c_{\b} (\n) \rangle ~=~ (\Delta \bh_{\a}, \Delta \bh_{\b}) \,.
\ee
The proof of the second equality is given in \cite{Dhurandhar2024}. The $\chi^2$ discriminator is then given by:
\be
\chi^2 ~=~ \Delta c_{\a} ~[C^{-1}]^{\a \b}~ \Delta c_{\b} \,.
\label{final}
\ee
By choosing an orthonormal basis of $\S$, one can explicitly show that in Gaussian noise, the $\chi^2$ defined in Eq. (\ref{final}) is the sum of squares of independent standard normal variables and hence the so defined statistic is $\chi^2$ distributed with $p$ degrees of freedom \citep{DGGB2017}.
\par

To extract maximum benefit from the generic $\chi^2$, it is intuitively clear that the $\bh_\a$ should be chosen in the direction of the glitches, so that they project heavily on $\S$.

\subsection{The underlying mathematical structure}

Recall that the space of data trains over a time segment $[0, T]$ is the Hilbert space $\D = L_2 ([0, T], \mu)$, where $\mu$ is a measure defined conveniently in the Fourier domain by $d \mu = df / S_h (f)$, with $S_h (f)$ being the one-sided noise PSD. The $\chi^2$ discriminator is defined so that its value for the signal is zero, and for Gaussian noise, it has a $\chi^2$ distribution with some number of degrees of freedom. In practice, this is of the order of a few tens to a hundred for a CBC search. In principle, $\D$ is infinite-dimensional because it is a space of functions. However, actually, since the data are sampled at some finite rate, it becomes a finite-dimensional space but with a large number of dimensions; typically, a data train contains points of the order of $\sim 10^6$ or more.
\par

Consider a single waveform $\bh$. Now choose a finite-dimensional subspace $\S (\bh) \subset \D$ of dimension $p$ normal to $\bh$, defined by:
\be
\S (\bh)   = \{\x \in \D ~|~ (\x, \bh) = 0 \} \,.
\ee
 Note that we are selecting vectors $\x$ from a plethora of such vectors so that they form a $p$ dimensional vector space. Then we claim that the $\chi^2$ pertaining to the waveform $\bh$ is just the square of the $L_2$ norm of an arbitrary data vector $\x$ projected onto $\S$ (we drop the $\bh$ in order to avoid clutter). Specifically, we decompose the data vector $\x \in \D$ as,
\be
\x = \x_{\S} + \x_{\S^\perp} \,,
\ee
where $\S^{\perp}$ is the orthogonal complement of $\S$ in $\D$ defined as:
\be
\S^{\perp} = \{\x \in \D ~| ~(\x, \y) = 0, ~~~~ \forall ~~\y \in \S \} \,.
\ee
We may write $\D$ as a direct sum of $\S$ and $\S^{\perp}$, that is, $\D = \S \oplus \S^{\perp}$.
The quantities $\x_{\S}$ and $\x_{\S^\perp}$ are projections of $\x$ into the subspaces $\S$ and $\S^{\perp}$, respectively. 

 Then the statistic $\chi^2$ is just,
\be
\chi^{2} (\x) = \| \x_{\S} \|^2 \,.
\ee
 Now choose any orthonormal basis of $\S$ say $\e_{\a},~~\a = 1, 2, ..., p$ so that $(\e_{\a}, \e_{\b}) = \delta_{\a \b}$, where $\delta_{\a \b}$ is the Kronecker delta. We easily verify the following properties:

\begin{enumerate}

\item For a general data vector $\x \in \D$, we have:
\be
\chi^2 (\x) = \| \x_{\S} \|^2 = \sum_{\a = 1}^p |(\x, \e_{\a})|^2 \,. 
\ee   

\item Clearly, $\chi^2 (\bh) = 0$, because $\bh$ is orthogonal to $\S$ and so the projection of $\bh$ into the subspace $\S$ is zero or $\bh_{\S} = 0$. 

\item Now let us take the noise $\n$ to be Gaussian with zero mean. Therefore, we obtain the following:
\be
\chi^2 (\n) = \| \n_{\S} \|^2 = \sum_{\a = 1}^p |(\n, \e_{\a})|^2 \,.
\ee
Observe that the random variables $(\n, \e_{\a})$ are independent and Gaussian, with mean zero and variance unity because $\langle (\e_{\a}, \n) (\n, \e_{\b}) \rangle = (\e_{\a}, \e_{\b}) = \delta_{\a \b}$ (see \cite{Dhurandhar2024} for proof). Thus $\chi^2 (\n) $ has a $\chi^2$ distribution with $p$ degrees of freedom. 
\end{enumerate}

The $\chi^2$ statistic so defined, satisfies the essential criteria of being zero on the signal and is distributed  $\chi^2$ for Gaussian noise. We call this the {\it unified} $\chi^2$ or the {\it generic} $\chi^2$. 
\vsss

One can think of the $p$-dimensional  subspace $\S$ as "attached" to each point of the signal manifold - we, therefore, have the structure of a vector bundle. The unified $\chi^2$ is a function on the base manifold, which is the signal manifold or the parameter space we had denoted by $\P$. Although, in principle, there is an enormous choice in selecting $\S$ and, therefore, the $\chi^2$ discriminator, physical and practical considerations limit the choice. Here, we mention two such important considerations:

\begin{itemize}

\item The normal spaces $\S$ chosen must be such that the projection of the glitches on $\S$ is large. Note that a glitch occurring in a data train is a vector in $\D$, denoted by $\bg$. Specifically, if $\bg$ is a glitch, then as before decomposing $\bg = \bg_{\S} + \bg_{\S^\perp}$, we must have $\chi^2 (\bg) = \chi^2 (\bg_{\S}) \gg p$. This will distinguish the glitch from the signal and Gaussian noise. Since this is not a straightforward proposition, special constructions are required to select the subspaces with this property. Glitches shaped like sine-Gaussians are ubiquitous in detector data, and an optimal $\chi^2$ to rule out such glitches has been constructed \cite{Joshi_2021}. Our analysis shows a lot of freedom in selecting $\S$, which should be advantageous in constructing better $\chi^2$ discriminators.

\item It is desirable that the computational cost of evaluating the $\chi^2$ is not too large - that is, the discriminator is computationally efficient. 
\end{itemize}

\subsection{The traditional $\chi^2$ discriminator}
\label{subsec:trad_chisqr}

The traditional method, due to Bruce Allen \cite{Allen2005}, involves partitioning the frequency band $[f_l, f_u]$ into $p$ sub-bands with the following properties: The frequency band $[\fl, \fu]$ is decided by the bandwidth in which the detector operates and the signal in which it has appreciable power. We form the subintervals $I_\a = [f_{\a - 1}, f_\a],~ \a = 1, 2, ..., p$, such that $f_0 = \fl$ and $f_p = \fu$ and there is equal signal power in the sub-bands. Explicitly for the binary inspiral signal (Newtonian waveform), we have $\th (f) = N_h f^{-7/6}$ where $N_h$ is a normalisation constant. We then normalise the signal by setting:
\bea
\langle \bh, \bh \rangle &=& 4 \int_{\fl}^{\fu}~ \frac{|\th (f)|^2}{S_h (f)}~ df \,, \no \\
                     &=& 4 N_h^2\int_{\fl}^{\fu}~ \frac{df}{f^{7/3} S_h (f)} ~\equiv~ 1 \,,
\eea
by choosing $N_h$ appropriately.

The $f_1, f_2, ..., f_{p-1}$ are so chosen that the above integral over $I_\a$ equals $1/p$, that is,
\be
4 \int_{f_{\a- 1}}^{f_\a}~ \frac{|{\tilde h} (f)|^2}{S_h (f)}~ df = \frac{1}{p}, ~~~~~~~~~~~k = 1,2, ..., p \,.
\ee
Given these definitions, we can state the Bruce Allen $\chi^2$ statistic in the following way. In the Fourier domain, we define the functions $\th_\a (f)$ as:
\bea
\th_\a (f) &=& \th (f) ,~~~~ f \in I_k \no \\
        &=& 0 ~~~~~{\rm otherwise} \,.
\eea
Then the quantities $\th_\a$ possess the following properties:
\be
\th = \sum_{\a = 1}^p \th_\a, ~~~~~~ \langle \th_\a, \th_\b \rangle = \frac{1}{p} \delta_{\a \b}, ~~~~~~ \langle \th_\a, \th \rangle =
\frac{1}{p} \,.
\label{eq:Allen_1}
\ee
We have omitted the argument $f$ to avoid clutter (and there is a slight abuse of notation for the scalar product). We can describe the above procedure in geometric language in vector form. We have the correspondence $\th (f) \longrightarrow \bh$ and $\th_\a (f) \longrightarrow \bh_\a$ where we have represented functions by vectors. For a template (normalised) $\bh$ and a data vector $\x$, the correlation is $c = (\x, \bh)$. We assume that the signal matches the template perfectly. The Bruce Allen test consists of splitting the template vector $\bh$ into $p$ frequency bands equal in power and checking for consistency. This translates Eq. (\ref{eq:Allen_1}) to:
\be
\bh = \sum_{\a = 1}^p \bh_\a, ~~~~~~~~~ (\bh_\a, \bh_\b) = \frac{1}{p} \delta_{\a \b}, ~~~~1 ~\leq~ \a, ~\b ~\leq p \,. 
\ee
Then from Eq. (\ref{eq:dhalpha}), we have $\Dh_{\a} = \bh_{\a}  - \frac{\bh}{p}$ and $\Delta c_{\a} = (\x, \Dh_\a)$. 
\vss

We observe that $(\Dh_{\a}, \bh) = 0$ for each $\a$ so that the subspace spanned by $\Dh_\a$ is orthogonal to $\bh$. This is precisely $\S$. We further see that $\sum_{\a = 1}^{p} \Dh_{\a} = 0$  so that the dimension of $\S$ is less than $p$. Since this is the only linear relation $\dim (\S) = p - 1$. Further we compute that,
$(\Dh_{\a}, \Dh_{\b}) = - \frac{1}{p^2}$ for $\a \neq \b $, implying that the $\Dh_\a$ are not orthogonal.
\vsss

We can however exhibit an orthonormal basis of $\S$: 
\bea
\e_1 &=& \sqrt{\frac{p}{2}} \left(\Dh_1 - \Dh_2 \right) \,, \no \\
\e_2 &=& \sqrt{\frac{p}{6}} \left(\Dh_1 + \Dh_2 - 2 \Dh_3 \right) \,, \no \\
\vdots &&     \no \\
\e_{p - 1} &=& \sqrt{\frac{p}{p(p - 1)}} \left(\Dh_1 + \Dh_2 + ... + \Dh_{p - 1} - (p - 1) \Dh_p \right) \no
\eea      
Thus, we have explicitly established that the traditional $\chi^2$ (i) has $p - 1$ degrees of freedom because it is the sum of the squares of $p - 1$ standard independent Gaussian random variables, and (ii) it is also a unified $\chi^2$.

\section{The optimised $\chi^2$ for sine-Gaussians  and blip glitches}
\label{sec:blip}

The construction described here is general and applicable to any family of glitches that can either be modelled or to a collection of samples, preferably of common morphology picked out from the data \cite{GBDC25}. Here, we  apply our formalism to the family of glitches, which are shaped like sine-Gaussians. These are the so-called blip glitches. The reason for this choice is that empirically, a large subset of noise transients in gravitational-wave strain data of LIGO and Virgo detectors has been found to project strongly on sine-Gaussians, including the types that trigger CBC templates~\cite{DGGB2017}.  In order to improve the sensitivity of CBC searches, we specifically target these glitches, which are a major source of reduced sensitivity. However, our algorithm can be straightforwardly adapted to replace sine-Gaussians with any other relevant glitch morphology.
\par

Below we detail the steps in obtaining an optimal $\chi^2$ (see \cite{Joshi_2021} for details).

\begin{itemize}
 \item Sample the parameter space of sine-Gaussians densely so that the sample is representative of the blip glitches. We call this subspace as $\VG$ (subscript $\G$ for glitches). Here the metric defined on the parameter space of glitches helps in uniformly sampling the parameter space.  
 
 \item Since the subspace $\S$ should be orthogonal to the trigger template $\bh$, we remove the component parallel to $\bh$ from each of the sample vectors spanning $\VG$. Thus if $\v \in \VG$, then we define $\v_\perp = \v - (\v, \bh) \bh$. By construction, the $\v_\perp$ are orthogonal to $\bh$. The space spanned by these clipped vectors, namely $\v_\perp$, we call $\Vp$.  
 
 \item In practice, $\VG$ or equivalently $\Vp$, contains a large number of vectors, few hundreds to thousands. If we used all these vectors to construct $\S$, its dimension would be large as well; these would be the number of degrees of freedom of the $\chi^2$, leading to high computational cost. Therefore we apply the Singular Value Decomposition (SVD) to the vectors of $\Vp$ to obtain a subspace of $\Vp$, that is the best fit low dimensional approximation to $\Vp$. 
 
 The best lower-dimensional approximation to a vector space of $n$ dimensions is defined as follows: given $n$ linearly independent vectors that span a vector space 
 $\V$, the best $k$-dimensional, $k < n$, approximation to $\V$ is the subspace $\W$, such that the sum of squares of the norm of the projections of all the $n$ vectors onto $\W$ is maximum.   
 
 How does one construct $\W$? This can be achieved by using the Eckart-Young-Mirsky theorem \citep{EY1936}. First one applies the SVD to the vectors that span $\V$. Then, one chooses the right singular vectors corresponding to the largest $k$ singular values, then the subspace spanned by these vectors is precisely $\W$. The dimension $k$ is chosen by applying a cut-off to obtain the desired level of approximation. The subspace $\S$ of $\Vp$ is generated in this way - it is the best $p$ dimensional approximation to $\Vp$. We also have the desired property that $\bh$ is orthogonal to $\S$ since $\S \subset \Vp$.
\end{itemize}

We start by sampling the space of sine-Gaussians in order to obtain $\VG$.

\subsection{The sine-Gaussian waveform and the metric}
\label{subsec:sine_Gauss}

\begin{figure}[htbp]
\begin{center}
\includegraphics[width=0.6\textwidth]{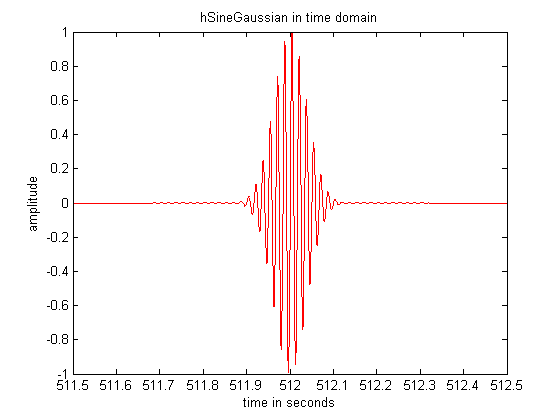}
\caption{The figure depicts a typical sine-Gaussian.}
\label{fig:sine_gaussian}
\end{center}
\end{figure}

\begin{figure}[htbp]
\begin{center}
\includegraphics[width=0.49\textwidth]{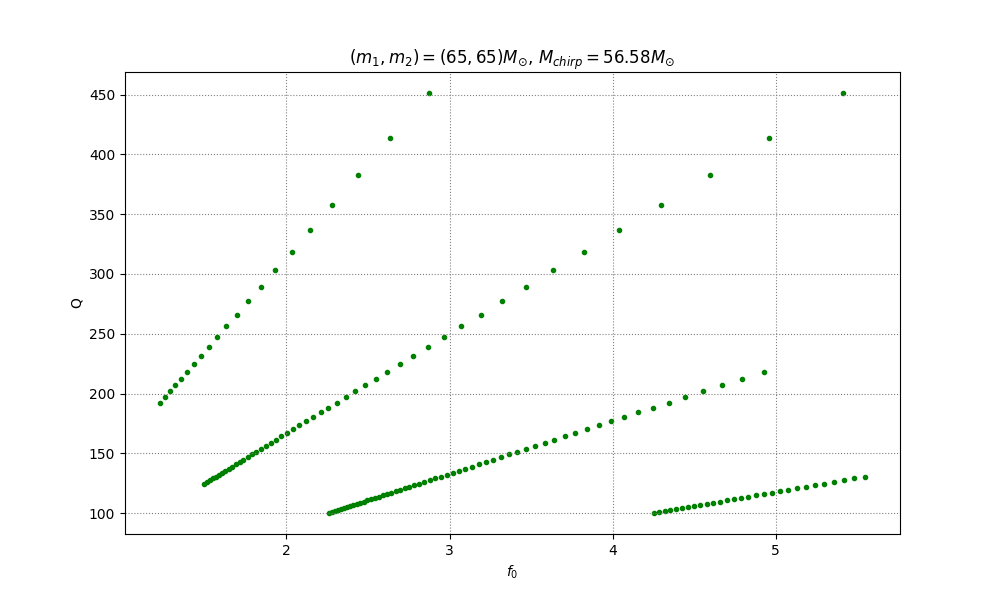}
\includegraphics[width=0.49\textwidth]{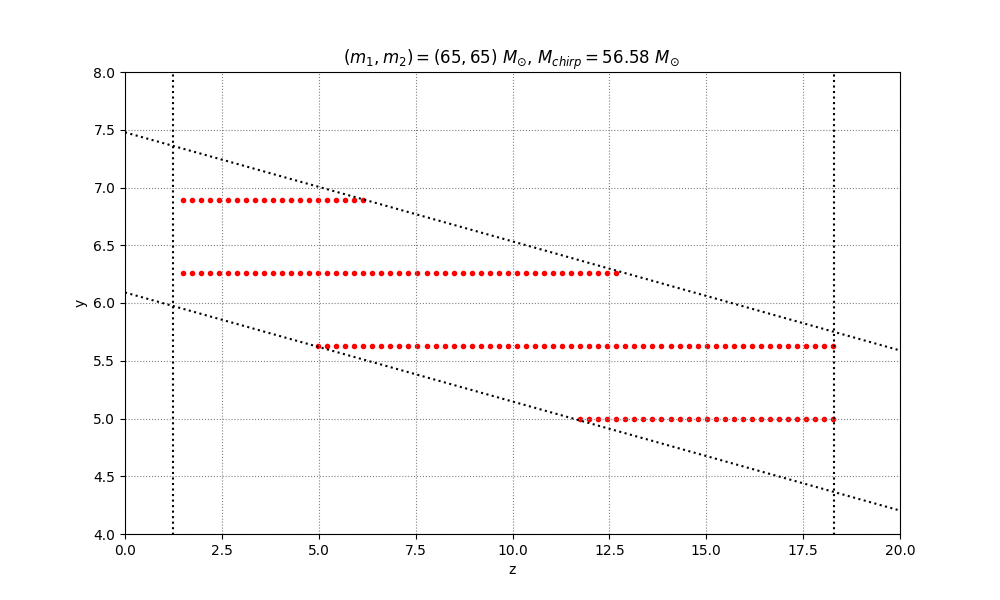}
\caption{The figure on the left displays the sampling of the $(f_0, Q)$ parameter space for $2 < Q < 8$ and $100 {\rm Hz} < f_0 < 500 {\rm Hz}$. The corresponding plot in the $(z-y)$ plane is shown on the right. The binary star masses are $m_1 = m_2 = 65 \Msun$ and the chirp mass $\Mc = 56.6 \Msun$. The match between any two adjacent sine-Gaussians is $80 \% $}
\label{fig:sampling}
\end{center}
\end{figure}

We begin by describing a sine-Gaussian waveform. Fig. \ref{fig:sine_gaussian} depicts the general form of this type of glitch.
We now proceed to design an optimal $\chi^2$ for blip glitches. It is therefore important to select appropriate vectors in the parameter space occupied by the glitches.
A recent work  \cite{Joshi_2021} demonstrates how transient bursts represented by sine-Gaussian waveform can be vetoed with the help of an optimal $\chi^2$ from the GW data. Since the blip glitches are also known to have a time-domain morphology similar to the sine-Gaussian waveforms, we use these waveforms as the vectors to generate $\VG$ and further construct the optimized sine-Gaussian $\chi^2$ - the Opt $\chi^2$. To allow for an arbitrary phase in the noise transient, we use a complex-valued sine-Gaussian waveform, which is more convenient.  In the time domain the sine-Gaussian waveform can be defined as,
\begin{equation}
\begin{aligned}
\psi(t;t_0,f_0,Q) =   & A\exp{\bigg(-\frac{4\pi^2f_0^2}{Q^2}(t-t_0)^2\bigg)} \\
& \times \exp{[-i2\pi f_0(t-t_0)]} \,. 
\end{aligned}
\end{equation}
where $t_0$ is the central time, $f_0$ the central frequency, 
$Q$ the quality factor and $A$ the amplitude. The parameters describing the glitches apart from the amplitude are $t_0,f_0$ and $Q$ and we will consider these three parameters to define the parameter space of sine-Gaussians. We have dropped the phase $\phi_0$ because the match is defined as the absolute value of the scalar product \cite{chatterji_thesis}, and so it does not appear anywhere in the computations - in effect, we have put it to zero. In order to populate the parameter space, we require a metric. This is achieved by considering two neighboring normalised 
sine-Gaussian waveforms, $\psi_1(f;t_0,f_0,Q)$ and
$\psi_2(f;t_0+dt_0, f_0+d f_0,Q+d Q)$. A metric may then  be introduced on this space as a map from the differences in the parameters of these waveforms to the fractional change in their match:
\begin{equation}
\begin{aligned}
ds^2 = & \bigg(\frac{4 \pi^2 f_0^2}{Q^2}\bigg)dt_0^2+\bigg(\frac{2+Q^2}{4f_0^2}\bigg)df_0^2+\bigg(\frac{1}{2Q^2}\bigg)dQ^2 \\
       & -\bigg(\frac{1}{2 f_0 Q}\bigg)df_0 dQ. 
\end{aligned}
\label{metric}
\end{equation}
The above metric (\ref{metric}) can be reduced to its diagonal form by using the transformations,
\begin{equation}
\omega_o=2\pi f_0, ~~~~~\nu=\frac{\omega_o}{Q}\,.
\end{equation}
In the new coordinates, $(t_0,\omega_0,Q)$, the  metric takes the form:
\begin{equation}
    ds^2=\nu^2 dt_0^2 + \frac{1}{4\nu^2} d\omega_0^2 + \frac{1}{2 \nu^2} d\nu^2.
    \label{intmid_metric}
\end{equation}
In comparison to the metric in Ref. \cite{Joshi_2021}, this metric has no $\omega_0^2$ term multiplying $dt_0^2$. This results from accounting for the arbitrary phase of the sine-Gaussian waveform. As mentioned in Ref. \cite{Bose_2016}, a CBC template is triggered with a time lag $t_d$ after the occurrence of a glitch, i.e., after $t_0$. The time $t_d$ is given by \cite{Bose_2016}, which for the parameters considered is essentially $\tau_0$, where, 
\begin{equation}
    \tau_0=\frac{5}{256\pi f_0}(\pi \Mc f_0)^{-5/3}.
\end{equation}
Here, the quantity $\Mc$ is called the chirp mass $\Mc = (\mu M^{2/3})^{3/5}$, where $\mu$ and $M$ are the reduced and the total mass of the binary system, respectively, and $\tau_0$ is called the Newtonian chirp time. Because of this relation between $t_0$ and $\omega_0$, the metric given in Eq.(\ref{metric}) is reduced to one in two parameters. In the new parameters,
\begin{equation}
z=(\omega_0 \Mc)^{-5/3}\quad {\rm and} \quad
y=\ln (\nu/\rm{rad/sec})\,,   
\end{equation}
the following metric results: 
\begin{equation}
ds^2=\Bigg[\frac{2^{-14/3}}{Q^2}+\frac{9Q^2}{100 z^2} \Bigg]dz^2+\frac{1}{2}dy^2.
\label{final_metric}
\end{equation}
For most of the parameter space we consider, the metric given in Eq.~(\ref{final_metric}) applies, and our sampling of the parameter space based on this metric is valid to a good approximation. Fig. \ref{fig:sampling} shows the sampling of the parameter space for a match of $80\%$ in the coordinates $(f_0, Q)$ and $(z, y)$.

\subsection{Obtaining $\S$: the degrees of freedom of the $\chi^2$}
\label{SVD}

Fig. \ref{fig:sampling} displays the sampling of the parameter space, which gives a collection of vectors which span the space $\VG$ on which the blip glitches are expected to have a substantial projection. However, as mentioned before, the SVD cannot be applied directly to the vectors in $\VG$. The input matrix, which we denote by $\bM$ needs to be first prepared. We briefly describe the procedure here since the full details can be found in \citep{Joshi_2021}. The following are the salient steps required:

\begin{itemize} 
\item The sine-Gaussians have central time $t_0 = 0$ and they need to be appropriately time-shifted with respect to the time of occurrence of the trigger.

\item  The vectors need to be {\em clipped} by subtracting out the component parallel to the template as described at the beginning of this section. The clipped and time-shifted sine-Gaussians span the subspace $\Vp$ of $\D$. 

\item The usual SVD algorithm uses the Euclidean scalar product. However, we have a weighted scalar product given by Eq. (\ref{eq:scalar}); a scalar product that is inversely weighted by the PSD $S_h (f)$. We therefore {\it whiten} each row vector by dividing each Fourier component of the row vector by $\sqrt{S_h (f)}$.
\end{itemize}

Now, we have a collection of vectors which we use to form the row vectors of $\bM$. This is the matrix to which the SVD algorithm is applied. After the SVD, we have to finally modify the output of the SVD, namely, the right singular vectors, which must then be {\it unwhitened}. This is done by multiplying by the factor $\sqrt{S_h (f)}$. The unwhitened singular vectors are orthonormal in the weighted scalar product.
\par

We now briefly describe the SVD algorithm \cite{NumericalRecipes}. The SVD decomposition of the input $M \times N$ matrix $\bM$ is written as a product of three matrices:
\begin{align}
\bM = \Ub ~\Sgm ~ \Vb^{\dagger}  \,,  
\end{align}
where $\Ub$ is the $M \times r$ matrix of left singular vectors, $\Sgm$ is an $r \times r$ square diagonal matrix consisting of singular values $\sigma_1, \sigma_2, ..., \sigma_r$. The singular values $\sigma_k, ~k = 1, 2, ,,,,r$ are arranged in descending order of magnitude. Finally, $\Vb^{\dagger}$ is the $r \times N$ matrix of right singular vectors. The matrix $\Vb^\dagger$ is the Hermitian conjugate of $\Vb$. Both left and right singular vectors are normalised and appear as column vectors in both matrices $\Ub$ and $\Vb$. Invoking  the Eckart-Young-Mirsky theorem we obtain the best $p$-dimensional approximation to $\Vp$, which  
states that the first $p$ singular vectors span the best-fit $p$-dimensional subspace of $\bM$. The cut-off $p$ is chosen by using Frobenius norm \citep{Matrix_comp-Frob} of the matrix $\bM$, namely, 
\begin{align}
\| \bM \|_F^2 = \sum_{i = 1}^M \sum_{j = 1}^N |a_{ij}|^2 \equiv \sum_{k = 1}^r \sigma_k^2 \,.
\end{align}
If we set the accuracy level to 90$\%$, then we choose $p < r$ such that $\sum_{k = 1}^p \sigma_k^2 ~\gtrsim ~0.9 ~\| \bM \|_F^2$. We obtain $\S$ as the linear span of the first $p$ right singular vectors $\v_1, \v_2, ..., \v_p$; they form an orthonormal basis of $\S$. This implies that the sum of squares of projections of the row vectors of $\bM$ on $\S$ add up to more than 90$\%$ of the full value of the sum of squares of all the vectors $\v_1, \v_2, ..., \v_M$. If a glitch vector is close to the span of the row vectors $\v_i, ~i = 1, 2, ..., p$, its square of the norm of its projection onto $\S$ will tend to be large, which will result in a large $\chi^2$. We have also succeeded in reducing the number of degrees of freedom of the $\chi^2$ to $p$, which is usually much less than the dimension of $\VG$.  Typically, for the ranges of parameters $f_0, ~Q, ~\Mc$, etc., considered here, the number of vectors spanning $\VG$ is about a few hundred 
while $p < 10$.  

\section{Results}
\label{sec:results}

\begin{figure}
\includegraphics[width=0.49\textwidth]{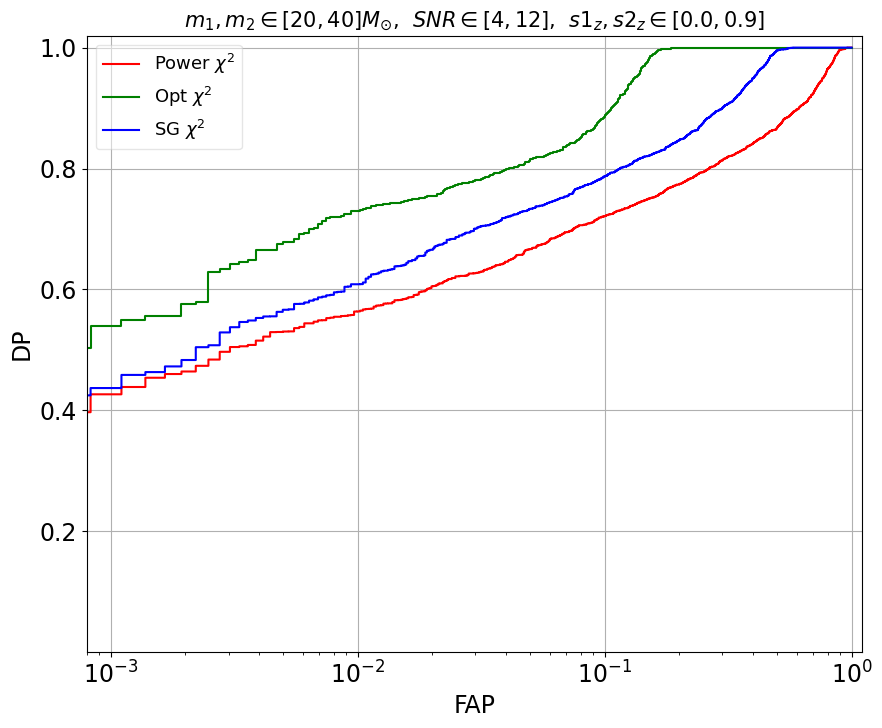}
\includegraphics[width=0.49\textwidth]{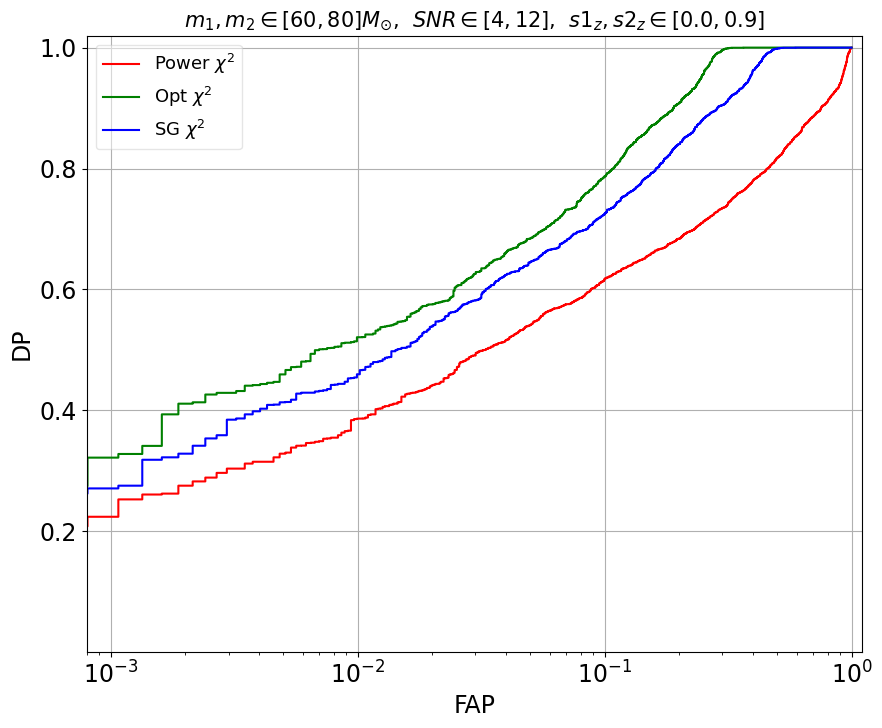}
\caption{The ROC curves for Opt, SG and traditional (power) $\chi^2$s were plotted for data trains containing blips and BBH signals for low (left figure) and high (right figure) mass bins. For each category, 2000 blips from LIGO O3 data and 2000 simulated CBC signals with component masses uniformly distributed in range 20 to 40  $M_{\odot}$ (left figure) and 60 to 80 $M_{\odot}$ (right figure). The aligned spins were in the range 0 to  0.9 and the SNRs in the range 4 to 12.}
\label{fig:rocs}
\end{figure}

 In this section, as a demonstration of our formalism, we evaluate the performance of the Opt $\chi^2$ as compared to the traditional (Power) and SG $\chi^2$, all of which differentiate blips from aligned spin black hole binary (BBH) signals. We selected real blips from LIGO's O3 run from both Hanford and Livingston detectors. We then considered a total of 4000 strain data segments (2000 each for each of the two mass bins (see below)) each of 16-second duration and containing a blip with a matched-filtering signal-to-noise ratio (SNR) between 4 and 12, as registered by the loudest BBH template. We further inserted BBH signals using the family of \textsc{IMRPhenomPv2} waveforms ~\cite{PhenomP}.  These simulated  signals also span the same SNR range, namely, 4 to 12 uniformly. We divide the signals into two bins based on their component masses: one bin consists of signals with component masses between 20 - 40 $M_{\odot}$ and the other between 60 - 80 $M_{\odot}$. In both cases, the aligned spins $s_{1z}$ and $s_{2z}$ are distributed uniformly in the range 0 to  0.9. \footnote{The dimensionless spins $s_{1z}. s_{2z}$ must be multiplied by the factor $G M^2/c$ to obtain the  dimensions of angular momentum. Here $G \sim 6.67 \times 10^{-8}$ is Newton's gravitational constant in cgs units and $c \sim 3 \times 10^{10}$ cm/sec the speed of light.} The purpose of this division is to 
compare the performance of the Opt $\chi^2$ in the two mass bins and therefore understand which part of the BBH parameter space benefits more in sensitivity, which was reduced because of the glitches. As an illustration, we display the results in Fig. \ref{fig:rocs} for SNR in the range 4 to 12. We found that only 3 (complex) basis vectors for the construction of the Opt $\chi^2$ were required - this amounted to 6 real degrees of freedom for the statistic \cite{CBDJ2024}.
In order to compute the matched-filter SNRs of both blips and spin-aligned BBH signals, we used different CBC template banks for the two mass bins, and the same range of spins as the injections. The template banks were generated with a minimum mismatch of 0.97 using the stochastic bank code of \textsc{PyCBC} \cite{pycbc-software}. The same \textsc{IMRPhenomPv2} waveform was used for templates as the one for simulating the CBC injections. The signals were injected into 16 sec long data segments from O3 which were free of any astrophysical signals. 
\par

For gauging the performance of the different $\chi^2$s the receiver-operating-characteristic (ROC) curves are plotted in Fig. \ref{fig:rocs}. We compare our Opt $\chi^2$ with the traditional (Power) $\chi^2$ and SG $\chi^2$. As one can see from the ROC curves in Fig.\ref{fig:rocs}, the Opt $\chi^2$ performs better than the traditional and SG $\chi^2$ at all false alarm probabilities in both low-mass and high-mass ranges. As can be observed in these results, the Opt $\chi^2$ performs at least as good as or even better than the other discriminators. This is likely because it exploits the differences and similarities between blips and BBH waveforms quantitatively. These advances have positively impacted the CBC search sensitivity over the traditional and the SG $\chi^2$s.

\section{Concluding Remarks}
\label{sec:concl}

Here we have presented a general framework for generic $\chi^2$ discriminators. These discriminators have the mathematical structure of a vector bundle. Then, the $\chi^2$ statistic is just the $L_2$ norm of the data vector projected onto a subspace of the space of data vectors orthogonal to the trigger template. Since the trigger could occur anywhere in the parameter space, we are dealing with a collection of subspaces attached to each point of the signal manifold - a smooth choice leads to a smooth vector bundle. The $\chi^2$ can be viewed as a non-negative real valued function on sections of the vector bundle.   Apart from this mathematical elegant construction, the important practical application that emerges from this formulation is the enormous flexibility available in the $\chi^2$ discriminators. This flexibility could be exploited for {\it tuning} the $\chi^2$;  it can discriminate more decisively against frequently occurring glitches or glitches which pose difficulties to the search algorithms. 
\par

One such morphology is that of blip glitches which can be modelled as sine-Gaussians. We have therefore constructed a $\chi^2$ statistic that is optimally effective in discriminating BBH signals from sine-Gaussian glitches \cite{Joshi_2021,CBDJ2024}. In previous works signal-based $\chi^2$ discriminators have been formulated that have succeeded in identifying noise artifacts in the data (see \cite{CBDJ2024} and references therein). However, in the recent past, they have been found wanting in high-mass BBH searches. This awareness has led to new proposals for discriminators which would improve on BBH search sensitivities. For the first time, reference \cite{DGGB2017} evolved the appropriate mathematical formalism for geometrically understanding existing $\chi^2$ discriminators and constructing new ones. 
\par

There are several steps, which we summarise below, required to arrive at the final goal. We considered a family of sine-Gaussian waveforms with physically relevant parameters. We then uniformly sampled this family of sine-Gaussians with the help of a metric so that it is adequately represented. The space spanned by this set of sampled sine-Gaussians we have called $\VG$. It turns out that the number of sampled glitch vectors is too large, and consequently, the subspace $\VG$ spanned by them is high dimensional. A
low-dimensional approximation to $\VG$ is sought in order that the computational costs for the $\chi^2$ remain in control. The
best possible low-dimensional approximation to $\VG$ is obtained by invoking the Eckart-Young-Mirsky theorem, and this is
achieved with the help of the SVD algorithm. We ensure that the associated subspace obtained for the $\chi^2$ is
orthogonal to the trigger template by appropriately projecting out the components of the glitch vectors parallel to the
trigger template. The above steps lead to the required optimal $\chi^2$ discriminator - the Opt $\chi^2$. Our results on real data from the O3 run of the LIGO detectors show that the Opt $\chi^2$ performs significantly better than the other $\chi^2$s. For more details see \cite{CBDJ2024}. Current work in progress \cite{GBDC25} uses real blip glitches extracted from detector data, instead of the sine-Gaussian model to construct the $\chi^2$. Additionally, this work also targets more than one glitch morphologies.  

\section*{Acknowledgments}

I would like to dedicate this work to the memory of Prof. Tanuka Chattopadhyay, whose invaluable contributions to astronomy and astrophysics, unbounded enthusiasm, and personal interaction will be deeply missed. 
\par

I gratefully acknowledge Tathagata Ghosh for Figures 2 and 3 and Sunil Choudhary for Figures 5 and 6. Some of the results have used the data, software, and/or web tools obtained from the Gravitational Wave Open Science Center \citep{gw_open_data, gwosc}, a service of LIGO Laboratory, the LIGO Scientific Collaboration, and the Virgo Collaboration.   Some of the simulations for this work were carried out at the IUCAA computing facility Sarathi. 
\par

I thank Abisa Sinha Adhikary and Asis Chattopadhyay for the generous hospitality shown by them in Kolkata. 
\bibliography{references.bib}
\end{document}